\documentstyle[prb,preprint,aps]{revtex}
\begin{document}                
%
\title{Frictional drag between spatially separated two-dimensional electron 
gases mediated by virtual phonon exchange}
\author{Samvel M. Badalyan$^{*}$ and Ulrich R\"ossler}
\address{Institute for Theoretical Physics, University of Regensburg, D-93040 Regensburg, 
Germany\\ (*) and Department of Radiophysics, Yerevan State University, 375049 
Yerevan, Armenia}
\draft%
\maketitle
\begin{abstract}                
We have calculated the temperature dependence of the frictional drag between 
spatially separated quantum wells with parallel two-dimensional 
electron gases due to interlayer electron-electron interaction mediated by 
virtual exchange of acoustic phonons due to piezoelectric and deformation 
potential interaction. It is shown that the frictional 
drag is dominated by the piezoelectric coupling. According to our calculations the 
temperature dependence of the drag scattering rate divided by $T^2$ exhibits a 
pronounced peak which for the experimental situation and in agreement with
the finding of T. J. Gramila, et al.\, Phys. Rev. B {\bf 47}, (1993) 12957
is obtained at about $T\approx 2.1$ K. We ascribe the appearance of this peak to 
a change from small to large angle scattering in the virtual phonon exchange.
\end{abstract}
\narrowtext

\section{Introduction}               

Double layer two-dimensional (2D) carrier systems as realized in semiconductor 
nanostructures are of interest with respect to fundamental phenomena 
as fractional quantum Hall effect and superconductivity because of the 
interlayer carrier interaction \cite{eisen}. 
Depending on the width of the 
potential barrier that separates the two layers one has to distinguish the 
regime of quantum-mechanical coupling via tunneling for small barrier widths
from that of interlayer coupling mediated by (direct or effective) 
electron-electron interaction for sufficiently large barriers. 
In the former case, the frictional aspects  of superconductivity 
\cite{lazovik,shevchenko} and of the fractional quantum Hall effect 
\cite{shayegan,eisenstein}  have been studied.
In the latter case, when the large barrier allows independent control of 
electron conduction in the two layers, the frictional drag effect 
manifests itself when a current driven along one layer (layer $1$) induces, 
via momentum transfer, a drag voltage in the second layer (layer $2$) under 
conditions that no current flows in this well \cite{pogreb,price}. 
The drag effect has been studied experimentally for two 2D electron systems 
(2DES), two 2D hole systems (2DHS), but also in situations with layer $1$ 
$2$ being a 2DES or 2DHS, respectively 
\cite{solomon,laikhtmann,sivan,gramila91,gramila92,gramila,rubel95,rubel96,flensberg96}.
Recently the effect of a magnetic field perpendicular to the 
layers has been investigated \cite{hill,rubel97}. Theoretical work on
the frictional drag effect was devoted to elucidating the microscopic mechanism
of the momentum transfer \cite{maslov,zheng,nikitkov}. 
The calculated temperature dependence of the frictional drag due to direct
Coulomb electron-electron interaction shows strong dependence on 
the inter-layer separation $\Lambda$ \cite{tso,jauho} while in  the  experiment
by Gramila et al.\ \cite{gramila}, the temperature dependence of the 
observed frictional drag for separations at least up to $\Lambda\sim 50$ nm  
exhibits almost no dependence on $\Lambda$, and has been interpreted as 
being due to the exchange of virtual phonons. The frictional drag accomplished 
by the exchange of virtual phonons has been studied theoretically in 
\onlinecite{peeters}.
This work, however, because of its conceptual complexity requires involved 
numerical computations and does not lead to analytical expressions for the 
dependence of the 
frictional drag on the system parameters. Therefore, we present here an 
analytical derivation of the drag scattering rate for the particular case of 
asymmetric two-layer systems ($n_1\neq n_2$) coupled by virtual exchange of 
acoustic phonons including its temperature dependence.
We thus model the experimental situation investigated by T. J. Gramila, 
et al.\ \cite{gramila} and find by using the corresponding system 
parameters the characteristic $T$-dependence of the drag scattering rate in 
quantitative agreement with these experimental data. 

\section{Theoretical concept}

Our model as depicted in Fig.\ \ref{fg1} consists of two parallel electron 
sheets separated by a distance $\Lambda$ with electron densities $n_1$ and 
$n_2$ and the layer extensions $d_1$ and $d_2$.
According to experimental observations, a current $I_1$ driven along the layer
$1$ by applying a voltage $V_1$ leads by frictional drag to a voltage $V_2$ in
the layer $2$. The drag scattering rate in a double layered system can be 
defined as \cite{flensberg96}
\begin{equation}
{1\over \tau_{Drag}}={E_2\over \tau_1 E_1}
\label{eq1}
\end{equation}  
where $E_1,\,E_2$ are the electric fields connected with the voltages
$V_1,V_2$, and $\tau_1$ is the transport relaxation time which determines
via the mobility $\mu=e \tau_1/m^*$ the current in the layer $1$; $m^*$ is 
the electron effective mass.
Following \onlinecite{jauho}, we obtain by using the linearized Boltzmann 
equation for the drag scattering rate
\begin{eqnarray}
{1\over \tau^\Upsilon_{Drag}}&=&{\hbar^2\over 4m T n_1 L^2}
\sum_{\sigma_{1},\sigma_{2},\sigma_{1'}
\sigma_{2'}}\sum_{\vec{k}_{1},\vec{k}_{2},\vec{k}_{1'}\vec{k}_{2'}}
q_\perp^2 W^\Upsilon_{1,2\to 1',2'}
\nonumber\\
&\times& f_1 f_2(1-f_{1'})(1-f_{2'})   
\label{eq2}
\end{eqnarray} 
where $L$ is the normalization length, $\vec{k}$ the electron in-plane 
momentum, $\sigma$ the electron spin, $\vec{q_\perp}=\vec{k}-\vec{k'}$ the 
transferred momentum, $f$ the Fermi distribution function determined by 
temperature $T$ and Fermi energy $\varepsilon_F=\hbar^2k_F^2/2m^*$ with Fermi 
wave vector related to the areal electron concentration $n$, by $k_F^2=2\pi 
n$. $W^\Upsilon_{1,2\to 1',2'}$ is the transition probability of two 
electrons from the states $|1>=|\sigma_{1},\,\vec{k}_{1}>$ and 
$|2>=|\sigma_{2},\, \vec{k}_{2}>$ with energies $\varepsilon_1$ and 
$\varepsilon_2$, respectively, into the states $|1'>=|\sigma_{1'},\, 
\vec{k}_{1'}>$ and $|2'>=|\sigma_{2'},\, \vec{k}_{2'}> $ with energies 
$\varepsilon_{1'}$ and $\varepsilon_{2'}$. The index $\Upsilon$ refers to the
type of effective electron-electron interaction which causes this transition.

In this work we consider effective electron-electron interaction mediated by the 
exchange of virtual acoustic phonons related to the piezoelectric ($\Upsilon=$ PA) 
and deformation ($\Upsilon=$ DA) potential coupling.
Using Fermi's Golden Rule, $W^\Upsilon_{1,2\to 1',2'}$ can be written in the
form
\begin{equation}
W^\Upsilon_{1,2\to 1',2'}= {2\pi\over\hbar} \left| T^\Upsilon_{1,2\to 1',2'} 
\right|^2 
\delta(\varepsilon_1+\varepsilon_2-\varepsilon_{1'}-\varepsilon_{2'}) 
\label{eq3}   
\end{equation} 
where $ T^\Upsilon_{1,2\to 1',2'}$ is the transition matrix element.  
Phonon mediated electron-electron interaction appears in second order
perturbation theory with respect to the electron-phonon coupling. The 
transition matrix element of this process is visualized by the diagram in 
Fig.~\ref{fg2}. For intrasubband scattering the vertex part of this diagram 
corresponds to
\begin{equation}
\Gamma^\Upsilon (\vec{q})=\sqrt{B^\Upsilon (q)}\:
\delta_{\vec{k}_{1},\vec{k}_{1'}+\vec{q}_\perp}
\:\int dz \rho(z) \exp(i q_z z)
\label{eq4}
\end{equation} 
with the subband electron density $\rho(z)$ and $\vec{q}=(\vec{q}_\perp,q_z)$.
$B^\Upsilon(q)$ depends only on the absolute value of $\vec{q}\:$ \cite{levinson}
\begin{eqnarray}
&& B^{PA}(q)=B^{PA}_0 q^{-1},\, B^{PA}_0={\hbar(e\beta)^2\over 2\varrho s},
\label{eq5}
\\
&&B^{DA}(q)=B^{PA}_0 q,\, B^{DA}_0={\hbar\Xi^2\over 2\varrho s}.
\label{eq6}   
\end{eqnarray} 
Here we describe electron-phonon interaction in the framework of the  
isotropic model with $s$ being the longitudinal sound velocity, $\varrho$ the 
crystal mass density, $e\beta$ and $\Xi$ are the piezoelectric and deformation
potential constants, respectively, averaged over the directions of the vector 
$\vec{q}$ and the phonon polarizations \cite{zook,kogan}. 
We assume also that all elastic parameters of the sample are the same so that 
phonons are not reflected from the interfaces separating different materials
(for instance between GaAs and AlGaAs).
The phonon propagator is
\begin{equation}
D(q)={2\hbar^{-1}\omega_{q} \over (\omega^2+i/2\tau_{q})^2-\omega_{q}^2}
\label{eq7}   
\end{equation} 
with $\hbar \omega=\varepsilon_1-\varepsilon_{1'}$, $\omega_{q}=s q$ and a 
summation over phonon momenta $\vec{q}$ corresponds to the internal phonon 
line (the dashed line in the diagram). In the above expression of the phonon
propagator, $\tau_{q}$ is the lifetime of the intermediate phonon states with 
respect to all processes which are destroying these intermediate states. For 
virtual intermediate states there is no energy conservation at the vertices of 
the diagram, {\it i.e.} $\varepsilon_1+\varepsilon_2=
\varepsilon_{1'}+\varepsilon_{2'}$ can be different from $\hbar\omega$ and
we may assume infinite lifetime $\tau_{q}$ for the intermediate phonons when 
calculating the drag rate. 
Then all integrals with respect to the energy parameters in Eq.\ (\ref{eq2}) are 
to be understood in the sense of the principal value.
Taking the summation over intermediate momenta $\vec{q}$, we obtain 
for the squared-modulus of the transition matrix elements for the two
considered mechanisms
\begin{eqnarray}
&&\left| T^{PA}_{1,2\to 1',2'} \right|^2={1\over \bar{\tau}_D^{PA}} 
{\hbar^3\over 2m} {(2\pi \Lambda)^2\over L^4} |I^{PA}(\alpha,\Lambda)|^2,\:
\label{eq8}
\\
&&\left| T^{DA}_{1,2\to 1',2'} \right|^2={1\over \bar{\tau}_D^{DA}} 
{\hbar^3\over 2m} {(2\pi \Lambda)^2\over L^4} {1\over (2k_F\Lambda)^4}
|I^{DA}(\alpha,\Lambda)|^2
\label{eq9}
\end{eqnarray} 
where 
\begin{eqnarray}
&&|I^{PA}(\alpha,\Lambda)|^2={e^{-2\alpha\Lambda}\over(\alpha\Lambda)^2}\;
{\sinh^2(\alpha d/2)\over (\alpha d/2)^2\left[1+(\alpha d/2\pi)^2\right]^2},
\label{eq10}
\\
&&|I^{DA}(\alpha,\Lambda)|^2={e^{-2\alpha\Lambda}\over(\alpha\Lambda)^2}\;
{(\omega\Lambda/s)^4\;\sinh^2(\alpha d/2)\over 
(\alpha d/2)^2\left[1+(\alpha d/2\pi)^2\right]^2}
\label{eq11}
\end{eqnarray}
and $\alpha$ is defined generally as
$\alpha(q_\perp,\omega+i\xi)=(q_\perp^2-(\omega+i\xi)^2/s^2)^{1/2}$ where 
$\xi$ is the positive infinitesimal, and the branch cut for the square root 
is assumed to lie along the negative real axis. It is easy to check that with 
such definition of $\alpha$, the expressions of the form factors (\ref{eq10}) 
and (\ref{eq11}) are valid both for $q_\perp^2-\omega^2/s^2>0$ and
$q_\perp^2-\omega^2/s^2<0$. In calculating the form factors $I^{PA}$ and 
$I^{DA}$ we have assumed that electrons are localized in symmetric infinitely 
high quantum wells with the width $d$, thus having $\rho_1(z)=(2/d)\,
\sin(\pi z/d)^2$ and $\rho_2(z)=\rho_1(z+\Lambda)$ as explicit forms for the 
electron density functions in layers $1$ and $2$, respectively.
In Eqs.\ (\ref{eq8}) and (\ref{eq9}) we introduce also nominal scattering 
times
\begin{eqnarray}
&&{1\over \bar{\tau}_D^{PA}}={2m(e\beta)^4\over \pi^2\hbar^3 \varrho^2
s^4}\,{ms^2\over\varepsilon_F}
\approx {1\over 0.5\mbox{$\mu$ s}},
\label{eq12}
\\
&&
{1\over \bar{\tau}_D^{DA}}={m\Xi^4 (2k_F)^4\over 2\pi^2\hbar^3 \varrho^2
s^4}\,{ms^2\over\varepsilon_F}
\approx{1\over 1.4\mbox{$\mu$ s}}
\label{eq13}
\end{eqnarray} 
with the numerical values obtained for a GaAs quantum well with the electron 
concentration $n_1=n_2=n\approx 1.5\cdot 10^{11}$ cm$^{-2}$ as in 
\onlinecite{gramila}.
Substituting Eqs.\ (\ref{eq3}),(\ref{eq8}), and (\ref{eq9}) into 
Eq.\ (\ref{eq2}) and taking integrals over $\vec{k}_{2'}$ and over directions 
of the vector $\vec{k}_2$ by exploiting $\delta$-functions, we can represent 
the drag scattering rate in the form
\begin{equation}
{1\over \tau^\Upsilon_{Drag}}={1\over \bar{\tau}_D^{\Upsilon}}\: {T^2\over 
(2\hbar s k_F)^2 }\: {\cal F}^\Upsilon(T)
\label{eq14}   
\end{equation} 
where 
\begin{eqnarray}
{\cal F}^\Upsilon(T)={1\over\pi}\:{(2k_F\Lambda)^2\over T^3}\: 
p.v. \int_0^\infty\!\! d\varepsilon_1 \!
\int_0^\infty\!\! d\varepsilon_{1'} \!
\int^\infty_{\varepsilon_0(q_\perp)}\!\! d\varepsilon_2 \nonumber \\
\times
\int_0^{2k_F}\!\! {dq_\perp\over \sqrt{4k_F^2-q_\perp^2}}
\,\sqrt{\varepsilon_{q_\perp}\over \varepsilon_2 - \varepsilon_0(q_\perp)}
\:|I^{\Upsilon}(\alpha,\Lambda)|^2
\nonumber \\
\times
f(\varepsilon_1) f(\varepsilon_2)\left(1-f(\varepsilon_{1'})\right)
\left(1-f(\varepsilon_1-\varepsilon_{1'}+\varepsilon_2)\right).   
\label{eq15}   
\end{eqnarray} 
In this equation $\varepsilon_{q_\perp}=\hbar^2 q_\perp^2/(2m^*)$ and the 
limiting energy
$\varepsilon_0(q_\perp)=(\hbar\omega-\varepsilon_{q_\perp})^2/
(4\varepsilon_{q_\perp})$ 
is obtained from the energy and momentum conservation laws. Now making use of 
the identity
\begin{equation}
f(x)\left(1-f(x+y)\right)={f(x)-f(x+y)\over 1-\exp(-y)}
\label{eq16}   
\end{equation} 
and taking integrals in Eq.\ (\ref{eq15}) with respect to the energy 
parameters, we reduce the expression for ${\cal F}^\Upsilon(T)$ to 
a one-dimensional integral of the form
\begin{eqnarray}
{\cal F}^{\Upsilon}(T)={\sqrt{\pi x_F}\over2}\:\gamma^2\, 
\int_0^1 {dx\over \sqrt{1-x}} {\beta^{\Upsilon}(x)\over\sinh^2{(\gamma\sqrt{x}/2)}}
\nonumber \\
\times
\left({PolyLog}({\frac{1}{2}},-{e^{-x_{+}}})
-{PolyLog}({\frac{1}{2}},-{e^{-x_{-}}}) \right)
\label{eq17}   
\end{eqnarray} 
where $PolyLog[\nu, z]$ gives the polylogarithm function of the order $\nu$ 
\cite{abramovic}
\begin{equation}
PolyLog[\nu, z]\equiv Li_\nu(z)=
{z\over \Gamma(\nu)} \int_0^\infty dt\, {t^{\nu-1}\over e^t-z}
\label{eq18} 
\end{equation} 
and
\begin{eqnarray}
&&x_{\pm}= x_F\left(\left({\gamma\over 4 x_F}\pm \sqrt{x}\right)^2-1\right),\:
\gamma={2\hbar s k_F\over T},\\ &&x_F={\varepsilon_F\over T},\quad
\beta^{\Upsilon}(x)=\cases{1,\quad \mbox{for PA} \cr {x^2},
\quad \mbox{for DA}}.
\label{eq19}   
\end{eqnarray} 
For the special GaAs system with $n\approx 1.5\cdot 10^{11}$ cm$^{-2}$ 
we have $\gamma\approx (7.6$ K)$/T$ and $x_F\approx (62.4$ $\mbox{K})/T$.
One can see from Eq.\ (\ref{eq17}) that in the framework of the adopted 
approximation when the lifetime of the intermediate phonon states is 
infinite, {\it i.e.} $\tau_{q}^{-1}=0$, the drag scattering rate, Eq.\ (\ref{eq14}), is 
independent of the inter-layer spacing $\Lambda$. This means that our
theory refers to the experimental situation when $\Lambda$ is smaller than the
phonon mean free path, $\Lambda_0=s\tau_q$, associated with the finite phonon
lifetime. For $\Lambda\gtrsim\Lambda_0$, we have to consider the finite $\tau_q$
in the phonon propagator, which would lead to an exponential dependence of the 
scattering probability of $W\propto\exp(-\Lambda/\Lambda_0)$.

\section{Results and discussion}

Using Eqs.\ (\ref{eq14}) and (\ref{eq17}) we plot the temperature dependence 
of the drag scattering rate divided by $T^2$ for identical layers in a symmetric
GaAs/AlGaAs double quantum well structure as in \onlinecite{gramila} coupled  both 
by PA- and DA-phonon mediated electron-electron interaction (Fig.\ \ref{fg3}). 
It is seen from this figure that the contribution of the DA interaction is smaller 
(approximately by one order of magnitude) than that of the PA interaction.
This is because the scattering probability due to long-range piezoelectric 
interaction includes an additional factor $q^{-4}$ with respect to the 
short-range deformation potential interaction. This difference is strongly 
pronounced at low temperatures \cite{chin} where the Pauli exclusion principle
restricts electron-phonon scattering processes in a Fermi gas to those with 
small angles.

Our calculations correspond to the experimental situation of Ref.\ \onlinecite{gramila} 
where the drag rate was measured for samples with the inter-layer spacing 
$\Lambda=17.5, \; 22.5\;$, and $50$ nm with the electron concentration 
$n=1.5 \cdot 10^{11}$ cm$^{-2}$ in a temperature range from $1$ to $7$ K.
Starting from the lower edge of this interval, where the drag rate is small
because it is dominated by small angle scattering connected  with a small  
momentum transfer, the drag rate increases with increasing temperature. 
Towards the upper edge of the interval, large angle scattering becomes 
dominant which reduces the increase of the drag scattering rate 
in temperature.
This gives rise to a peak of the function ${1/(\tau^\Upsilon_{Drag}}\,T^2)$ 
both for drag caused by the exchange of DA- and PA-phonons at approximately 
the same temperatures as shown in Fig.\ \ref{fg3}. The maximum drag rate 
of about $1$ s$^{-1}$ K$^{-2}$ due to combined effect of DA- and PA-phonons is
obtained at a temperature near $T_p\approx 2.1$ K in good agreement 
with the experimental results for the samples with the inter-layer spacing 
$\Lambda=17.5, \; 22.5,\; 50$ nm \cite{gramila}.
Our theory is not applicable for much larger interlayer spacing (as {\it e.g.}
for the sample with $\Lambda=500$ nm in Ref.\ \onlinecite{gramila}).
We believe that in this sample due to other 
scattering mechanisms, the intermediate virtual phonon states which 
effectively realize electron-electron interaction, are  destroyed and this 
leads to the exponential suppression of the drag scattering rate observed in 
the experiment.

In Fig.\ \ref{fg4} we present the temperature dependence of the drag 
scattering rate for a symmetric double quantum well system as in 
Fig.\ \ref{fg3} but for different values of the matched electron density
$n$ (The second line from the top ($n=1.5 \cdot 10^{11}$ cm$^{-2}$) is 
identical with solid line in Fig.\ \ref{fg3}.). With increasing $n$ the
peak structure becomes broader while the maximum value decreases and
the position $T_p$ shifts to higher temperature. This shift follows
approximately the relation $T_p\propto k_F\propto\sqrt{n}$ (see the 
inset of Fig.\ \ref{fg4}) which 
supports our findings that the peak position is related to the energy
$\hbar s k_F$ which separates the regions of small ($\hbar\omega\ll
\hbar s k_F$) and large ($\hbar\omega\sim \hbar s k_F$) angle 
scattering. We have also studied the drag scattering rate as in 
Fig.\ \ref{fg3} with fixed density $n_1=1.5 \cdot 10^{11}$ cm$^{-2}$
but changing density $n_2$ at different temperatures. A peak occurs
always at matched densities for which the momentum transfer is most 
efficient. It is seen for $T=6.5$ K and most clearly pronounced at 
$T=2.1$ K while for $T=1.1$ K it is only a small feature on an 
otherwise monotonous function of $n_2$. This strikingly different
behavior for $T=1.1$ K as compared to $T=2.1$ and $6.5$ K which is
similar to the experimental finding in Refs.\ \onlinecite{gramila,rubel95},
can be attributed to the changing dominance of small and large angle
scattering with temperature and $n_2$.

In conclusion, we have calculated the drag scattering rate between 
spatially separated electron layers due to the virtual phonon exchange 
via DA and PA mechanisms We have shown that the frictional drag due 
to PA coupling is dominating. Assuming infinite lifetime of the intermediate 
virtual phonon states, the frictional drag does not depend on the interlayer 
separation.
According to our calculations the temperature dependence of the drag 
scattering rate divided by $T^2$ exhibits a pronounced peak both at 
matched electron density of two layers and at temperatures where transition
from small to large angle scattering occurs.
These results are in agreement with the finding of J. Gramila, et al.\ \cite{gramila}.

\acknowledgments
We acknowledge support of this work by the Deutsche Forschungsgemeinshaft
(GRK: "Komplexit\"at in Festk\"orpern"). SMB acknowledges also support 
from the Deutscher Akademischer Austauschdienst (DAAD) and the US Civilian 
Research \& Development Foundation for the Independent States of the Former Soviet 
Union (CRDF, Award No. 375100).

\begin{figure}  
\caption{Schematic view of the double-layer system with the characterizing 
parameters (described in the text).}
\label{fg1}
\end{figure}
\begin{figure}  
\caption{Diagram of the effective electron-electron interaction between 
layers $1$ and $2$ as mediated by exchange of virtual acoustic phonons.}
\label{fg2}
\end{figure}
\begin{figure}  
\caption{Drag rate divided by the squared absolute temperature versus
temperature calculated for the symmetric GaAs/AlGaAs double layer
system of Ref.\ 13. The contributions of deformation potential
(DA, dotted line) and piezoelectric coupling (PA, dashed line) are 
shown together with the sum (solid line).}
\label{fg3}
\end{figure}
\begin{figure}
\caption{Drag rate (sum of DA and PA) divided by the squared 
temperature as a function of temperature plotted for the symmetric 
GaAs/AlGaAs double-layer system as in Fig.\ 3 but with different electron 
concentrations. Inset shows the peak positions (squares) corresponding
to drag rate data from this figure as a function of square root of the 
matched electron densities. Solid line is a guide to the eye.}
\label{fg4}
\end{figure}
\begin{figure}  
\caption{Drag rate (sum of DA and PA) divided by the squared 
temperature for fixed first layer electron density $n_1$ as a function
of the second layer electron density $n_2$ at different temperatures.}
\label{fg5}
\end{figure}

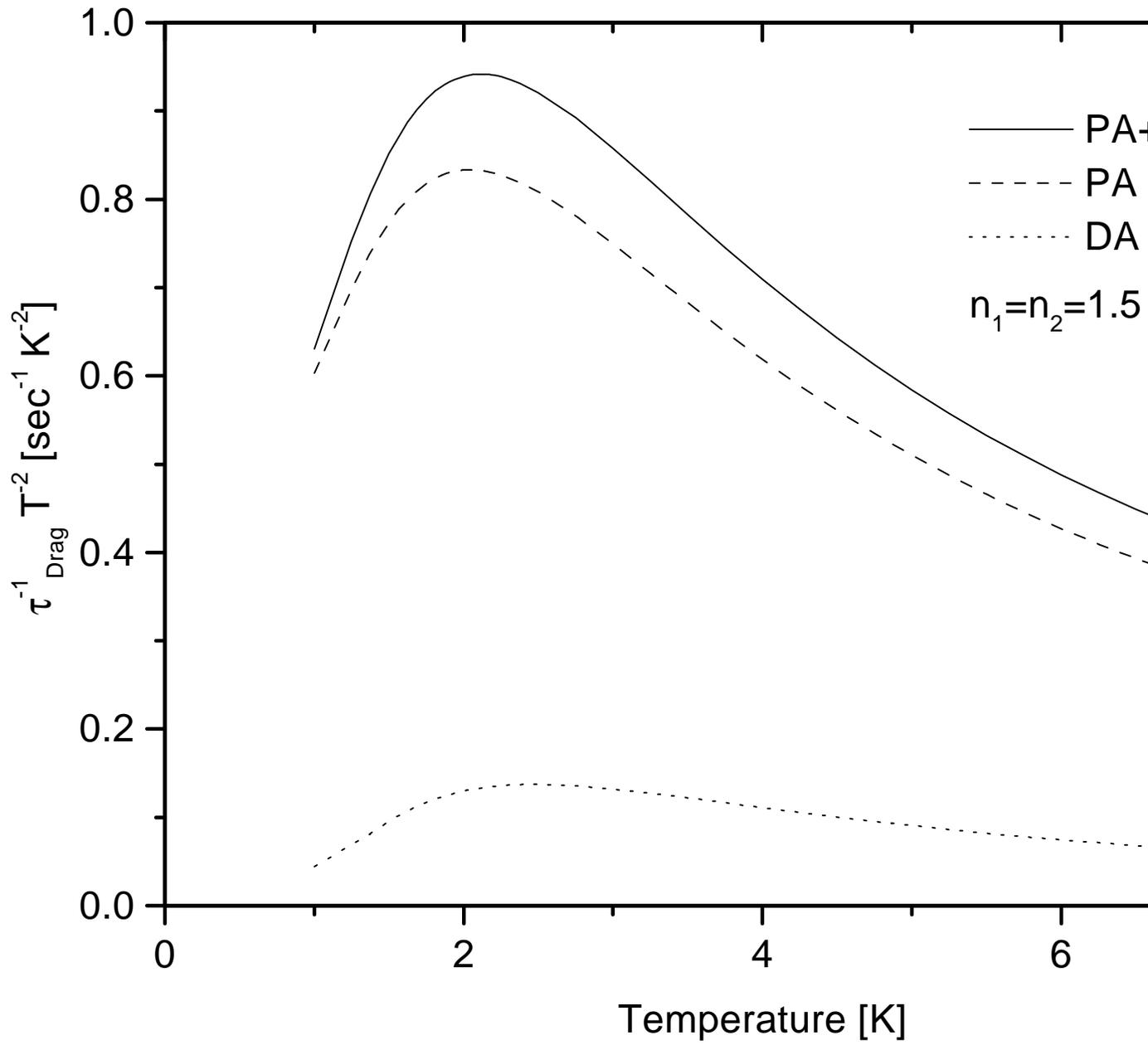
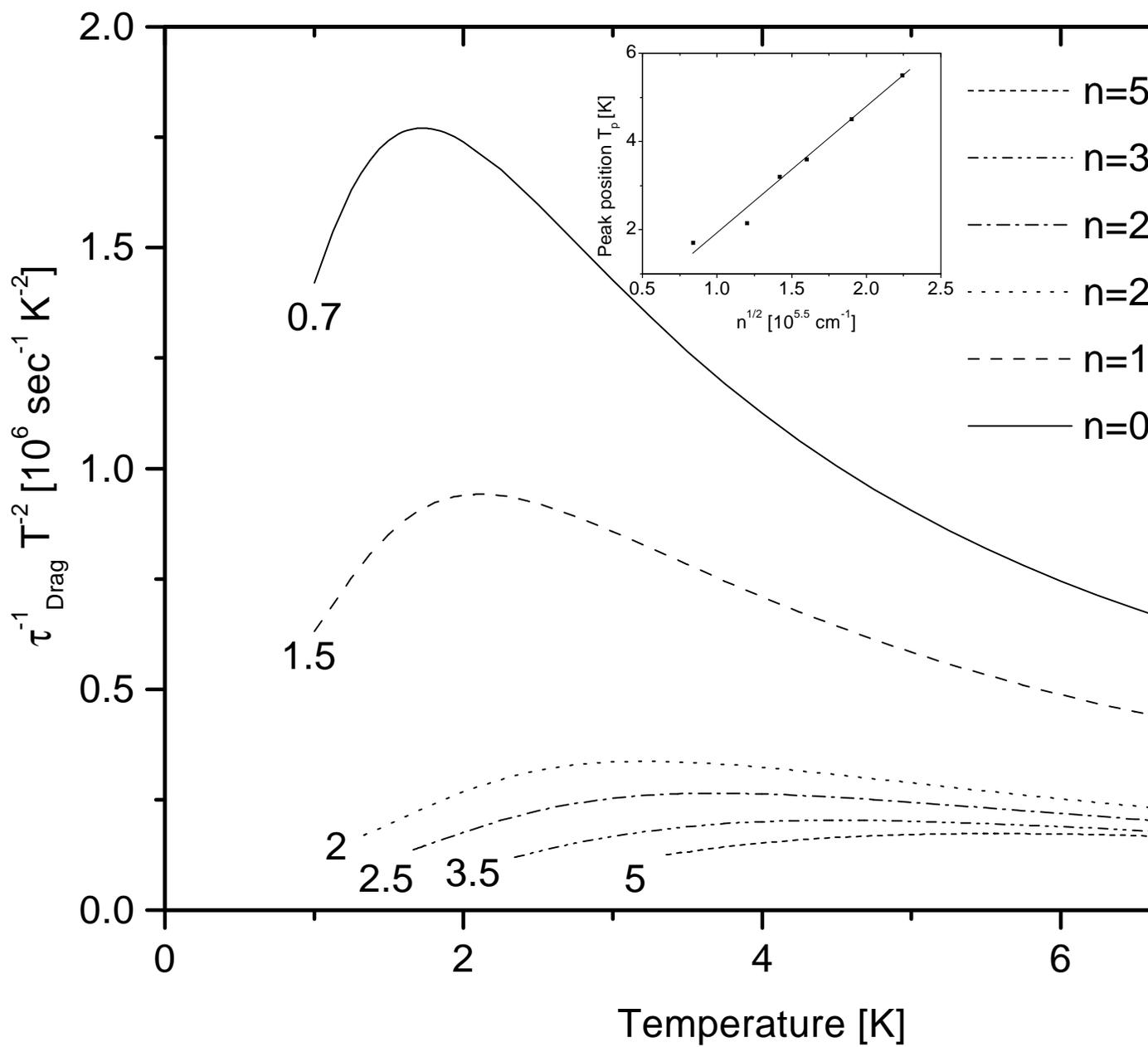
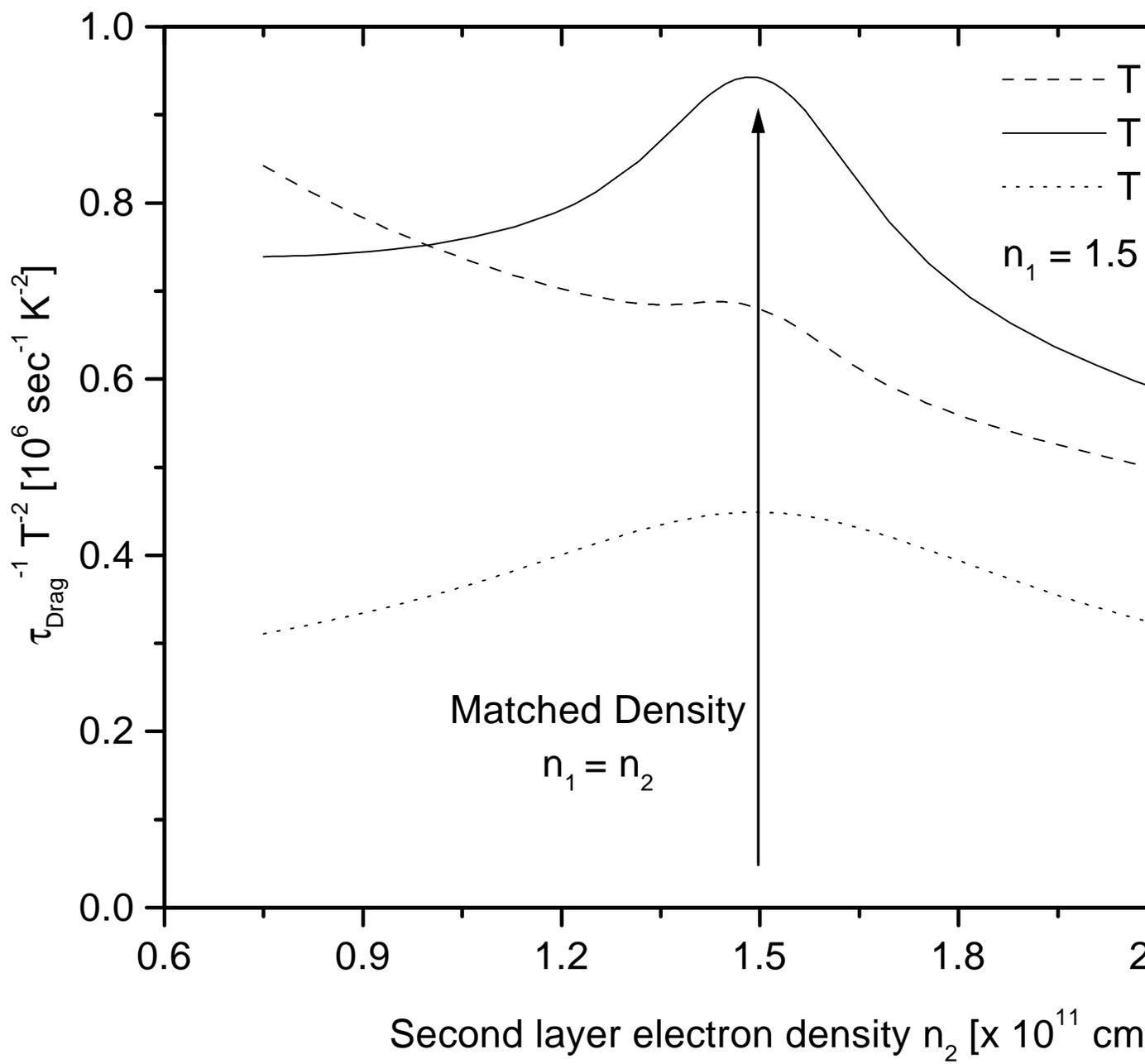
\end{document}